\documentclass[
reprint,
superscriptaddress,
amsmath,amssymb,
aps,
prb,
]{revtex4-2}
\usepackage[english]{babel}
\usepackage{amsmath}
\usepackage{amssymb}
\usepackage[bookmarks=true,colorlinks,citecolor=blue,linkcolor=blue,allcolors=blue]{hyperref}
\usepackage[capitalise]{cleveref}
\usepackage{physics}
\usepackage[toc,page]{appendix}
\usepackage{graphicx}

\begin{document}

\title{Time-dependent Neural Galerkin Method for Quantum Dynamics}

\author{Alessandro Sinibaldi}
\email{alessandro.sinibaldi@epfl.ch}
\affiliation{Institute of Physics, \'{E}cole Polytechnique F\'{e}d\'{e}rale de Lausanne (EPFL), CH-1015 Lausanne, Switzerland}
\affiliation{Center for Quantum Science and Engineering, EPFL, Lausanne, Switzerland}

\author{Douglas Hendry}
\affiliation{Institute of Physics, \'{E}cole Polytechnique F\'{e}d\'{e}rale de Lausanne (EPFL), CH-1015 Lausanne, Switzerland}
\affiliation{Center for Quantum Science and Engineering, EPFL, Lausanne, Switzerland}

\author{Filippo Vicentini}
\affiliation{CPHT, CNRS, École Polytechnique, Institut Polytechnique de Paris, 91120 Palaiseau, France}
\affiliation{Coll\`ege de France, Universit\'e PSL, 11 place Marcelin Berthelot, 75005 Paris, France}

\author{Giuseppe Carleo}
\affiliation{Institute of Physics, \'{E}cole Polytechnique F\'{e}d\'{e}rale de Lausanne (EPFL), CH-1015 Lausanne, Switzerland}
\affiliation{Center for Quantum Science and Engineering, EPFL, Lausanne, Switzerland}

\date{\today}

\begin{abstract}
We introduce a classical computational method for quantum dynamics that relies on a global-in-time variational principle.
Unlike conventional time-stepping approaches, our scheme computes the entire state trajectory over a finite time window by minimizing a loss function that enforces the Schr\"odinger's equation. 
The variational state is parametrized with a Galerkin-inspired ansatz based on a time-dependent linear combination of time-independent Neural Quantum States. 
This structure is particularly well-suited for exploring long-time dynamics and enables bounding the error with the exact evolution via the global loss function.
We showcase the method by simulating global quantum quenches in the paradigmatic Transverse-Field Ising model in both 1D and 2D, uncovering signatures of ergodicity breaking and absence of thermalization in two dimensions.
Overall, our method is competitive compared to state-of-the-art time-dependent variational approaches, while unlocking previously inaccessible dynamical regimes of strongly interacting quantum systems.
\end{abstract}

\maketitle
\paragraph*{Introduction --}
The major limitation to exact calculations in quantum many-body physics is the exponential growth of the Hilbert space, which makes systems of more than a handful of particles practically inaccessible to brute-force approaches.
Many-body variational methods are a powerful tool to circumvent this issue: instead of manipulating intractably large quantum states, a compressed representation relying on a smaller set of variational parameters can be used. 
Several classes of variational states have been employed to study quantum dynamics, including various incarnations of Tensor Network wave functions~\cite{white1992density,daley_time-dependent_2004}, as well as Neural Quantum States (NQS)~\cite{carleo2017solving}.

In the context of time-dependent NQS, conventional approaches to variational dynamics rely on the explicit integration of the Schr\"odinger's equation to obtain a different variational state at each time-step. 
This is realized either by means of a stochastic implementation of the time-dependent variational principle (TDVP)~\cite{carleo_localization_2012,carleo_light-cone_2014,carleo2017solving} or by projecting the exactly evolved state at each time in the variational manifold~\cite{Donatella2023Infidelity,sinibaldi2023unbiasing,nys2024ab,Gravina2024ptVMC}.
However, those schemes suffer from an accumulation of errors coming from the sequential propagation of the small time-step dynamics.

To circumvent those issues, we investigate a global-in-time variational principle that optimizes the entire time trajectory at once, departing from the sequential time-stepping paradigm and directly yielding the whole time-dependent solution to the initial value problem.
The principle consists of minimizing the deviations from the Schr\"odinger's solution at every time simultaneously, while the variational ansatz must be able to parametrize the quantum state throughout the entire dynamics.
Analogous approaches are employed in the field of Partial Differential Equation integration~\cite{Lagaris1998ANNODE,Sirignano2018ANNPDE}, as ways to circumvent the requirement to construct a discrete mesh, or in the field of Physics-Informed Neural Networks (PINNs) to solve a complex differential equation 
\cite{Rackauckas2020UDE,Cai2021PINNReview}.
Previous efforts to extend these concepts to quantum mechanical systems~\cite{wang2021spacetime} have achieved limited success when compared to conventional approaches. 
The numerical studies have been constrained to small-scale systems and imaginary-time evolution, falling short of demonstrating practical advantages.
While the exact reason remains unclear, we identify three major problems in the literature: the loss functions used (i) did not fully respect all gauge invariances of the Hilbert space, (ii) underfit the initial condition~\cite{Wang2021PINN_IVP_pathology,Krishnapriyan2021PINNFailureModes} and (iii) the deep neural architectures employed before suffer from a spectral bias, making them incapable of learning high-frequency components~\cite{Wang2020WhyPINNFail,Wang2021PINNSpectralBias}.

In order to move past such issues, we develop a global-in-time algorithm combining ideas from Machine Learning literature and physical requirements on the wave function parametrization.
Our key contributions are twofold. 
First, we design a physically-motivated loss function that preserves the fundamental quantum mechanical requirements of norm and phase invariance, enabling stable and accurate optimization. 
Second, we introduce a Galerkin-inspired ansatz based on a time-dependent linear combination of time-independent NQS, for which the deviation from the exact dynamics can be bounded through the loss function. 
Our approach is especially tailored for studying fundamental questions in quantum many-body physics, such as thermalization~\cite{kaufman2016quantum,reimann2016typical,RevModPhys.91.021001,saha2024thermalization,maceira2024thermalization,andersen2024thermalization}, many-body localization~\cite{nandkishore2015many,smith2016many,choi2016exploring,RevModPhys.91.021001,sierant2024many}, and hydrodynamic behavior~\cite{sanz2014many,castro2016emergent,banks2019emergent}, where accurate long-time dynamics is essential but traditionally difficult to access.
We apply our framework to investigate quantum quench dynamics in the 1D and 2D Transverse Field Ising model, reaching competitive accuracy with state-of-the-art variational approaches and unveiling signatures of lack of thermalization in two dimensions. 

\begin{figure*}[t]
    \centering
\includegraphics[width=0.8\linewidth]{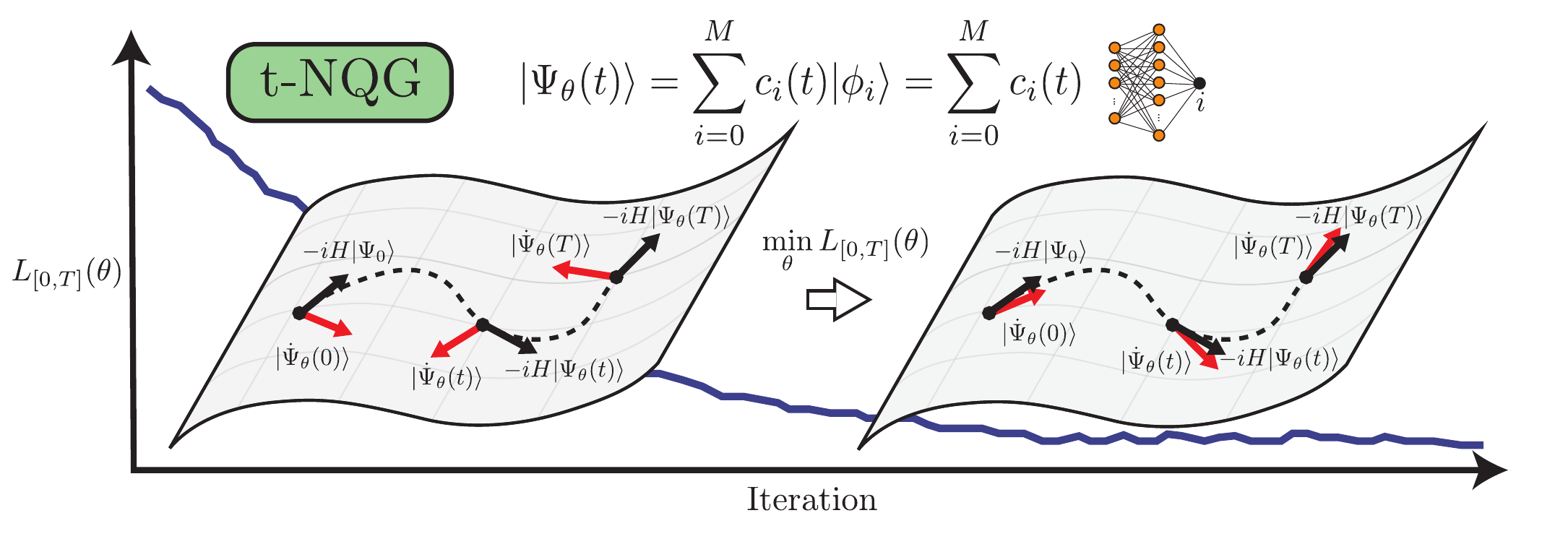}
    \caption{Sketch of the time-dependent Neural Quantum Galerkin (t-NQG) method for the simulation of quantum dynamics.
    The approach consists in minimizing the global loss function $L_{[0, T]}(\theta)$ in~\cref{eq:global_loss} matching $-i H\ket{\Psi_{\theta}(t)}$ (black arrow) and $\ket*{\dot{\Psi}_{\theta}(t)}$ (red arrow) at each time $t \in [0, T]$ in the subspace of the projector $P_{\perp \ket{\Psi_{\theta}(t)}} = 1 - \frac{|\Psi_{\theta}(t)\rangle\langle\Psi_{\theta}(t)|}{\langle\Psi_{\theta}(t)|\Psi_{\theta}(t)\rangle}$.
    The grey surface represents the variational manifold of the ansatz.
    The normalizations of the states are not indicated in the figure for simplicity.
    The ansatz consists of the linear combination of $M+1$ time-independent basis states $\ket{\phi_i}$ parametrized as Neural Quantum States (NQS) with time-dependent coefficients $c_i(t)$. 
    }
    \label{fig:artistic}
\end{figure*}

\paragraph*{Global-in-time variational principle --}
Our approach is based on a global dynamical variational principle that directly targets the entire time evolution, rather than evolving the state sequentially by integrating a local-in-time set of differential equations as in time-dependent Variational Monte Carlo (t-VMC)~\cite{carleo_localization_2012,carleo_light-cone_2014,carleo2017solving,Yuan2019theoryofvariational} or by performing projections like in projected t-VMC~\cite{gutierrez2022real,jonsson2018neural,medvidovic2021classical,Donatella2023Infidelity,sinibaldi2023unbiasing,Gravina2024ptVMC}.

Given a time-dependent quantum state $|\Psi(t)\rangle$ belonging to the Hilbert space $\mathcal{H}$, the Schrödinger's equation
\begin{equation}
\label{eq:schrodinger}
\frac{d}{dt}|\Psi(t)\rangle = -iH|\Psi(t)\rangle,
\end{equation}
determines the evolution of the state under the Hamiltonian $H$, which for simplicity we consider time-independent~\footnote{Our derivation would be equivalent in the case of a time-dependent Hamiltonian.}.

We encode the solution to the equation above with a differentiable time-dependent variational state $|\Psi_{\theta}(t)\rangle$ depending on a set of parameters $\theta$.
We stress the difference from \textit{established} time-dependent NQS approaches where the time-dependency is encoded in the parameters, as $\ket*{\Psi_{\theta(t)}}$.
The time-independent parameters $\theta$ that give a valid solution to~\cref{eq:schrodinger} can be determined by minimizing some distance between the left-hand side and the right-hand side of the equation at all times.
We start from the $L^2$ distance and additionally impose the physical requirements of norm~\footnote{By working with normalized ansätze, such as autoregressive NQS, we could forego this requirement. However, imposing this invariance comes at a negligible computational cost.} and phase invariance (see the Supplemental Material~\cite{suppmat} for a detailed derivation). 
This leads to the following time-local loss function, measuring the physically relevant deviations from the Schrödinger's dynamics,
\begin{equation}
\label{eq:loss}
\mathcal{L}(\ket{\Psi_{\theta}}) = \bigg|\!\bigg|  P_{\perp \ket{\Psi_{\theta}}}\bigg(\frac{|\dot{\Psi}_{\theta}\rangle}{\sqrt{\langle\Psi_{\theta}|\Psi_{\theta}\rangle}} + i H \frac{|\Psi_{\theta}\rangle}{\sqrt{\langle\Psi_{\theta}|\Psi_{\theta}\rangle}}\bigg)\bigg|\!\bigg| ^2, 
\end{equation}
where $P_{\perp \ket{\Psi_{\theta}}} = 1 - \frac{|\Psi_{\theta}\rangle\langle\Psi_{\theta}|}{\langle\Psi_{\theta}|\Psi_{\theta}\rangle}$. 
To keep the notation concise, we omit the explicit time dependence of the variational state.
The solution to the Schr\"odinger's equation in the time interval $[0,T]$ can be obtained by minimizing the integrated loss function 
\begin{equation}
\label{eq:global_loss}
L_{[0,T]}(\theta) = \frac{1}{T}\int_0^T dt\, \mathcal{L}(\ket{\Psi_{\theta}(t)}),
\end{equation}
assuming that the initial condition $\ket{\Psi_\theta(t=0)} = \ket{\Psi_0}$, where $\ket{\Psi_0}$ is the initial state of the dynamics, is fulfilled. 

The loss function~\cref{eq:global_loss} is positive semi-definite, $L\geq0$, and attains the minimum value of $L=0$ when $\ket{\Psi_\theta(t)}$ exactly satisfies the Schr\"odinger's equation~\cref{eq:schrodinger} at all times in the considered interval $[0, T]$. 
We stress that, as detailed in the Supplemental Material~\cite{suppmat}, this loss does not require the state $\ket{\Psi_{\theta}(t)}$ to have a unit norm, which means that, contrary to previous schemes featuring a global loss function~\cite{wang2021spacetime,PhysRevB.103.024304}, it is compatible with generally non-normalized variational parametrizations of the wave function.

To evaluate~\cref{eq:global_loss} in practice, we use an efficient Monte Carlo estimator (see the Supplemental Material~\cite{suppmat} for a detailed derivation)
\begin{equation}
\label{eq:global_loss_mc}
L_{[0,T]}(\theta) = \frac{1}{T}\int_0^T dt\, \mathbb{E}_{|\Psi_{\theta}(\sigma, t)|^2}[|\Bar{L}_\text{loc}(\sigma, t)|^2],
\end{equation}
where $\Bar{L}_\text{loc}(\sigma,t) = L_\text{loc}(\sigma,t) -\mathbb{E}_{|\Psi_{\theta}(\sigma,t)|^2}[L_\text{loc}(\sigma, t)]$ and $L_\text{loc}(\sigma,t)$ is a statistical estimator. 
In particular, the estimator corresponds to $L_\text{loc}(\sigma,t) = O_t(\sigma,t) + iE_\text{loc}(\sigma,t)$, namely the residual between the logarithmic time derivative $O_t(\sigma,t) = \partial_t\log\Psi_{\theta}(\sigma,t)$ and the local energy $E_\text{loc}(\sigma,t) = \langle\sigma|H|\Psi_{\theta}(t)\rangle/\Psi_{\theta}(\sigma,t)$.
Each term in the integral expression~\cref{eq:global_loss_mc} can be efficiently evaluated using standard Markov Chain Monte Carlo techniques by sampling configurations from $|\Psi_{\theta}(\sigma,t)|^2$. 
The time integration can be approximated using any scheme for the numerical computation of 1D integrals. 
In our calculations, we rely on the Simpson's $1/3$ rule~\cite{press1988numerical} with an odd number of equally spaced integration points. 
To find the minimum of~\cref{eq:global_loss_mc}, we resort to the standard Adam optimizer~\cite{kingma2017adam}.
As the convergence becomes increasingly challenging as $T$ is larger, we perform the optimization over smaller time sub-intervals and sequentially concatenate the solutions across adjacent time windows.
This procedure is explained more extensively in Appendix A. 

\paragraph*{Time-dependent Neural Quantum Galerkin --}
The global loss function $L_{[0,T]}(\theta)$ allows complete freedom for the choice of the time-dependent variational ansatz $|\Psi_{\theta}(t)\rangle$. 
One possible choice is to directly parametrize the wave function $\Psi(\sigma, t)$ as a function of the configurations $\sigma$ and the time $t$, as already proposed in~\cite{wang2021spacetime} by using a neural network architecture. 
However, such \textit{unstructured} approaches are known to lead to poor generalization beyond the time-interval $[0,T]$ considered by the loss function.

In this manuscript, we instead adopt an approach inspired by the Galerkin method~\cite{evans2022partial}, taking an ansatz which consists in a linear combination of $M+1$ time-independent basis states $|\phi_i\rangle$ with time-dependent coefficients $c_i(t)$,
\begin{equation}
\label{eq:ansatz}
|\Psi_{\theta}(t)\rangle =  \sum_{i=0}^M c_i(t)|\phi_i\rangle.
\end{equation}

In the previous expression, $\ket{\phi_0} \equiv \ket{\Psi_0}$ is the fixed initial state of the dynamics and $\{|\phi_i\rangle\}_{i=1}^M$ are variational states with parameters $\theta_i$, such that $\theta$ includes $\theta_i$ for $i=1, \ldots, M$. 
We remark that all the states in~\cref{eq:ansatz} are in general not normalized. 
The basis states $\{|\phi_i\rangle\}_{i=1}^M$ are taken to be Neural Quantum States (NQS)~\cite{carleo2017solving} with time-independent parameters, while the coefficients $\{c_i(t)\}_{i=0}^M$ can be parametrized via an expansion on a complete basis of 1D functions, for instance.
We name our method as \emph{time-dependent Neural Quantum Galerkin} (t-NQG). 
The t-NQG framework is sketched in~\cref{fig:artistic}. 

Any possible NQS architecture for standard Variational Monte Carlo can be used for t-NQG, including recurrent neural networks~\cite{hibat2020recurrent,roth2020iterative,hibat2021variational,hibat2022supplementing,wu2023tensor,ibarra2024autoregressive},
convolutional neural networks~\cite{liang2018solving,liang2018solving,roth2021group,roth2023high,fu2024lattice} and transformers~\cite{viteritti2023transformer_1d,viteritti2023transformer_2d,zhang2023transformer,sprague2024variational,lange2024transformer,rende2024mapping,rende2024simple}.
Provided the variational basis states are expressive enough, the specific ansatz in~\cref{eq:ansatz} can be made arbitrarily close to the exact solution of the Schr\"odinger's equation by increasing $M$. 
The optimal number of basis states typically scales linearly with the total evolution time $T$, more precisely $M \sim N \times T$ where $N$ is the system size. 
In the Supplemental Material~\cite{suppmat}, we show this scaling explicitly for a particular choice of the $\ket{\phi_i}$, corresponding to the basis states used in the coarse-grained approach of Ref.~\cite{PhysRevB.103.024304}.
In~\cite{suppmat} we also investigate numerically the role of the hyperparameters $M$ and $T$ in the optimization.
In Appendix E, we demonstrate that the algorithm can generalize beyond the time interval $[0, T]$ used for the training.

By minimizing~\cref{eq:global_loss} with the ansatz~\cref{eq:ansatz}, we obtain a variationally optimized set of basis states $\ket{\phi_i}$ and corresponding time-dependent coefficients $c_i(t)$.
Once these optimal states are determined, our linear ansatz naturally enables a refinement of the variational solution for the coefficients.
This is achieved by incorporating the trajectories $c_i(t)$ obtained via the time-dependent linear variational method with the fixed optimized basis states, as explained in Appendix C.
Furthermore, the structure of the ansatz provides a means to extrapolate the long-time limit of quantities such as expectation values of observables or the loss value itself, as detailed in Appendix D.

Finally, we remark that for the ansatz~\cref{eq:ansatz} the loss~\cref{eq:global_loss} represents a bound on the deviation from the exact solution of the dynamics, namely $|\!| e^{-i t H} | \Psi_0 \rangle - |\Psi_{\theta}(t) \rangle|\!| \leq t \sqrt{L_{[0, t]}}$.
Consequently, the error in the expectation value of any observable $O$ obeys 
\begin{equation}
|\delta O (t)| \leq |\!|O|\!|_2(2 t \sqrt{L_{[0,t]}} + t^2 L_{[0,t]}),
\end{equation} where $|\!|O|\!|_2$ indicates the operator 2-norm of $O$.
The derivation of these bounds is reported in the Supplemental Material~\cite{suppmat}. 

\begin{figure*}[t!]
    \centering
\includegraphics[width=1.0\linewidth]{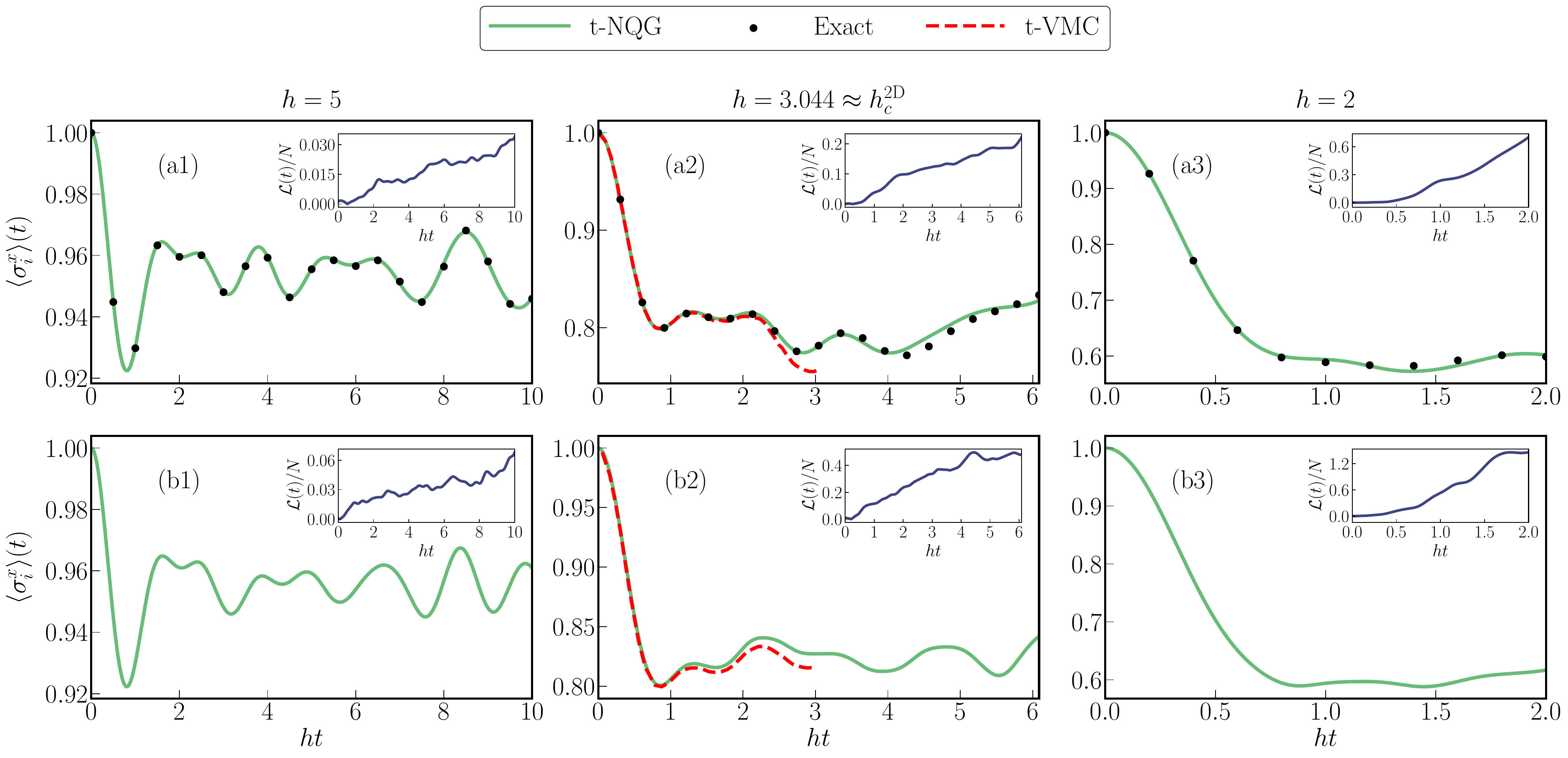}
    \caption{
    Time evolution of the transverse magnetization $\langle \sigma_i^x \rangle (t)$ following global quantum quenches in the TFI model on a $6 \times 6$ (upper panels) and $8 \times 8$ (lower panels) lattices.
    The system is quenched from the paramagnetically polarized initial state $|\Psi_0\rangle = \bigotimes_{i=1}^{N} |+\rangle_i$ to (a1–b1) the paramagnetic phase at $h = 5$, (a2–b2) the critical point at $h = 3.044 \approx h_c^{\mathrm{2D}}$, and (a3–b3) the ferromagnetic phase at $h = 2$.
    The basis states are represented by Restricted Boltzmann Machines (RBMs).
    For the $6 \times 6$ lattice, we employ $M = 18$ basis states and 512 Monte Carlo samples per integration point, while for the $8 \times 8$ lattice we use $M = 8$ basis states and 256 samples.
    The insets show the evolution of the loss function $\mathcal{L}(t) \equiv \mathcal{L}(|\Psi_{\theta}(t)\rangle)$, normalized with the system size $N$, as a measure of the variational accuracy.
    }
    \label{fig:2d_together}
\end{figure*}

\paragraph*{Results --}
To demonstrate the effectiveness of our approach, we consider the dynamics in the Transverse Field Ising (TFI) model with Hamiltonian
\begin{equation}
\label{eq:tfi}
H_{\text{TFI}} = -J\sum_{\langle i,j \rangle}\sigma^z_i\sigma^z_j - h\sum_i \sigma^x_i, 
\end{equation}
where $\sigma^{z, x}_i$ are the $z, x$ Pauli matrices acting on site $i$, $J$ is the coupling strength, $h$ is the transverse magnetic field and $\langle i,j \rangle$ indicates nearest-neighbor sites. 
Without loss of generality, we set $J=1$.
The TFI model exhibits a quantum phase transition in correspondence with the critical fields $h_c^{\text{1D}}=1$ for the 1D chain~\cite{mbeng2024quantum} and $h_c^{\text{2D}} \approx 3.044$~\cite{blote2002cluster} for the 2D square lattice, separating a ferromagnetic phase for $h<h_c$ from a paramagnetic phase for $h>h_c$.
It is a paradigmatic example where the interplay between the interaction and the field leads to rich dynamical behaviors.
We simulate the dynamics of global quenches in the 2D TFI model of $N$ spins with periodic boundary conditions.
In particular, we prepare the ground state of $H_{\text{TFI}}$ with $h=\infty$, namely the paramagnetically polarized state $|\Psi_0\rangle = \bigotimes_{i=1}^{N}|+\rangle_i$, and evolve it under $H_{\text{TFI}}$ for different values of $h$. 
This setup has been used as a benchmark in other variational calculations based on NQS~\cite{schmitt2020quantum,Gravina2024ptVMC}.
The time-independent basis states are encoded as complex-valued Restricted Boltzmann Machine (RBM) ansätze~\cite{carleo2017solving}, while the time-dependent coefficients $c_i(t)$ are expanded in a truncated Fourier basis (see Appendix B for details).

~\cref{fig:2d_together} shows the time evolution of the transverse magnetization following quenches in a $6 \times 6$ and $8 \times 8$ lattices. 
For the smaller system, where exact benchmark results are available, we observe that t-NQG accurately reproduces the quench dynamics in different phases up to fairly large times. 
For the more challenging quench at $h \approx h_c^{\text{2D}}$, we also compare with the state-of-the-art t-VMC simulation~\cite{schmitt2020quantum}, demonstrating that t-NQG is capable to reach higher precision and longer times. 
This improvement stems from the fact that t-VMC accumulates errors during the dynamics due to the local-in-time integration, while our method maintains high accuracy even at long times by directly targeting the entire trajectory.
While t-VMC can only access the information at the current time-step, t-NQG takes a global perspective on the full trajectory and can adjust the early dynamics to better accommodate later evolution.
The $h=2$ quench turns out to be arduous due to the more complicated optimizations, probably coming from the difficulty in learning ferromagnetic basis states.
This is evidenced by the loss function values during the dynamics, shown in the insets, which correlate with the actual error relative to the exact time evolution.
The loss function is normalized with $N$ to make it system size independent, since it is proportional to the energy variance which typically scales linearly with the number of degrees of freedom~\cite{wu2024variational}.
The time evolutions in the larger $8 \times 8$ lattice, where the exact dynamics is not accessible, are consistent with the trajectories obtained for the smaller system size.
Similarly, we observe that t-VMC~\cite{schmitt2020quantum} undershoots with respect to t-NQG for the critical quench.
We note that the normalized losses in the insets are of the same order as the ones for the $6 \times 6$, supporting the accuracy of the calculations for the larger lattice.
Once again, we remark increased complexity in simulating the $h=2$ quench, in line with the results for the smaller system.
In Appendix F, we report analogous results for a 1D lattice as a benchmark.

\begin{figure}[t]
    \centering
\includegraphics[width=1.0\linewidth]{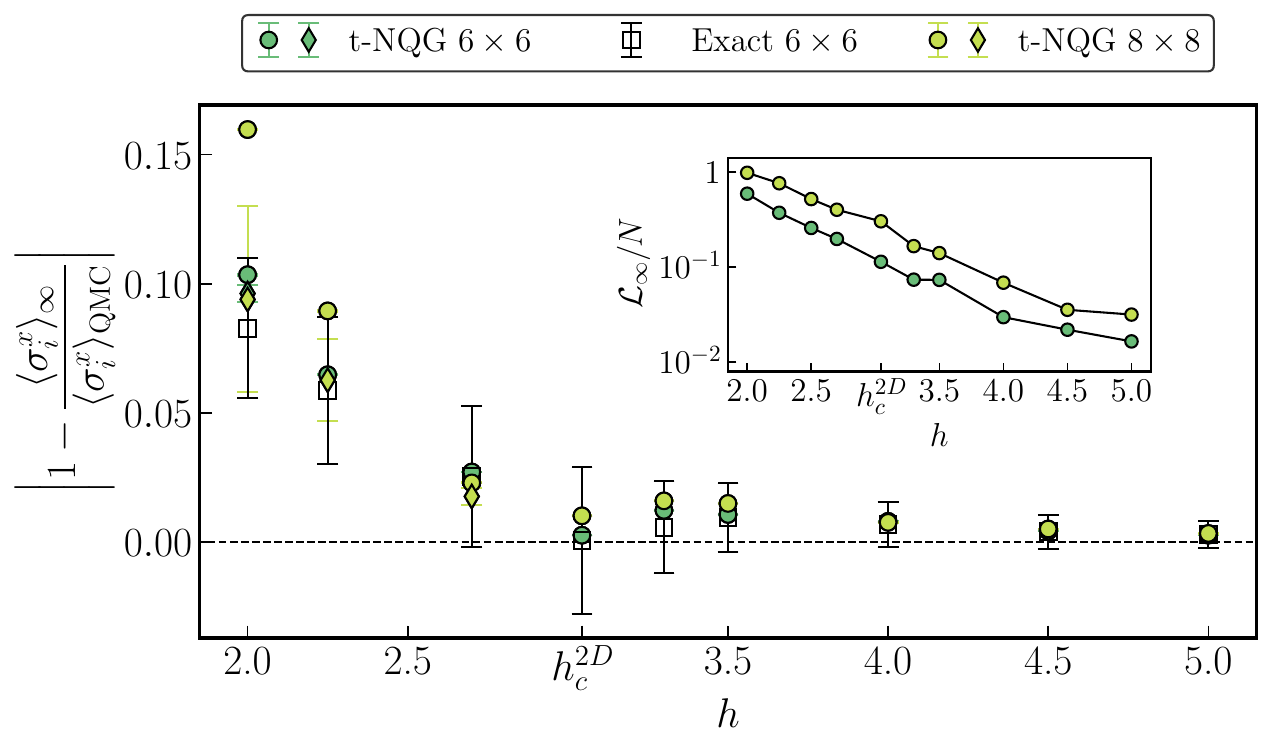}
    \caption{
    Relative deviation of the infinite-time transverse magnetization predicted by t-NQG, $\langle \sigma_i^x \rangle_{\infty}$, from its thermal value computed via Quantum Monte Carlo, $\langle \sigma_i^x \rangle_{\text{QMC}}$, for the $6 \times 6$ and $8 \times 8$ lattices across several quenches.
    Circle markers denote the t-NQG results, while diamonds indicate zero-loss extrapolations (shown only when they exhibit a distinguishable deviation) calculated as in Appendix G.
    The asymptotic exact values are also reported for the $6 \times 6$ system.
    Error bars on the long-time t-NQG data are assigned by repeating the calculation for 10 independent realizations, while those for the exact results arise from averaging over a finite time window.
    The inset shows the infinite-time loss function, $\mathcal{L}_{\infty} \equiv \lim_{t \rightarrow \infty} \mathcal{L}(|\Psi_{\theta}(t)\rangle)$, normalized by the system size $N$, which serves as a measure of the variational accuracy.
    }
    \label{fig:both_sizes}
\end{figure}

For the 2D lattices, we also compute the infinite-time value of the observable, as shown in Appendix D, for the different quenches.
We compare the result with the thermal expectation value $\langle O \rangle_{\mathrm{therm}} = \mathrm{Tr}[e^{-\beta_{\mathrm{eff}} H} O]/\mathrm{Tr}[e^{-\beta_{\mathrm{eff}} H}]$ to test the thermalization hypothesis~\cite{srednicki1994chaos,blas_test_2016}.
The effective inverse temperature $\beta_{\mathrm{eff}}$ is fixed by the conservation of energy condition, namely by solving the equation $\langle H \rangle_{\mathrm{therm}} = \langle \Psi_0 | H | \Psi_0 \rangle / \bra{\Psi_0} \ket{\Psi_0}$.
In the presence of ergodic dynamical behavior, the effective thermal average should coincide with the long-time dynamics, according to the Boltzmann prescription.
The thermal expectation values are computed by Quantum Monte Carlo (QMC) simulations based on the loop algorithm~\cite{todo_cluster_2001,albuquerque_alps_2007,bauer_alps_2011}.
The data are presented in~\cref{fig:both_sizes}. 
For the $6 \times 6$, we observe excellent agreement between the infinite-time t-NQG calculations and the extrapolations from the exact dynamics across all quenches considered.
This demonstrates that our method can faithfully extrapolate to the infinite-time limit while accessing only a portion of the finite-time evolution.
The inset illustrates that, although the variational accuracy -- quantified by the loss function at infinite time -- decreases for smaller $h$, the deviations from the exact results remain small.
This indicates that the long-time predictions of t-NQG are notably robust to the variational error, as suggested in the Supplemental Material~\cite{suppmat}.
The infinite-time loss remains of the same order for both system sizes, confirming that the accuracy for the $8 \times 8$ lattice is comparable to that of the smaller system, albeit slightly lower.
The variational error for the $8 \times 8$ can be systematically reduced by improving the convergence of the method~\cite{suppmat}.
Concerning the thermalization,~\cref{fig:both_sizes} reveals that for both lattice sizes the infinite-time value of the observable matches the effective thermal average for $h \gtrapprox  h_c^{\text{2D}}$, meaning that these quench dynamics are ergodic and thermalize in the long-time limit. 
In this regime, the $8 \times 8$ deviations are comparable to, or slightly larger than, those of the $6 \times 6$, likely reflecting the somewhat higher loss for the larger system.
For $h \lessapprox h_c^{\text{2D}}$, instead, the t-NQG predictions deviate significantly from the QMC calculations, with these deviations becoming more pronounced as the system size increases.
The zero-loss extrapolated values (see Appendix G) further support the robustness of the results obtained for the $8 \times 8$.
This behavior suggests a potential breakdown of ergodicity and thermalization, where the system becomes trapped in long-lived metastable states when driven far from equilibrium. 
This phenomenon is reminiscent of the behavior observed in interacting lattice bosons~\cite{carleo_localization_2012}.
We note that the deviation between the long-time dynamics and the thermal calculation could also arise from limitations in our finite-time window, as the t-NQG fit may not extend far enough to accurately capture the long-time behavior. 
A definitive assessment of this intriguing non-thermalization phenomenon in the quenched 2D TFI model would require additional systematic investigations across different system sizes and a more fine-grained exploration of the quench regimes.
To this end, enhancing the method with more expressive ansätze and higher-order optimizers~\cite{sorella1998green,sorella2005wave} would be key to minimizing the variational error of the simulations.

\paragraph*{Conclusion --}
In this work, we introduce a classical variational method for simulating the dynamics of many-body quantum systems. 
The scheme optimizes the entire quantum trajectory at once avoiding the accumulation of errors typical of time-stepping algorithms. 
Our approach relies on a global-in-time variational principle, in the form of a physically motivated loss function enforcing the Schr\"odinger's equation at each time, and employs a Galerkin-inspired ansatz based on Neural Quantum States (NQS).
This framework is particularly powerful for exploring the long-time limit of the dynamics and provides an effective bound on the error relative to the exact evolution.
We name our method \emph{time-dependent Neural Quantum Galerkin} (t-NQG).
We demonstrate the capabilities of t-NQG by simulating global quench dynamics in the 1D and 2D Transverse Field Ising model, achieving competitive performance with established time-dependent variational schemes.
In two dimensions, we unveil signatures of lack of ergodicity and thermalization when driving the system far from equilibrium. 

This work paves the way for leveraging NQS to study unexplored out-of-equilibrium phenomena in strongly-correlated quantum systems. Many extensions and applications could be envisaged. 
We mention that a straightforward extension would involve using more expressive deep neural network architectures as basis states, beyond the simple RBM employed here. 
Moreover, applications to benchmark noisy quantum computers are especially natural, extending the capabilities of other classical approaches typically limited to short time scales or one-dimensional geometries. 

\paragraph*{Data availability --}
The numerical simulations with the t-NQG method are based on NetKet~\cite{netket2,vicentini2022netket}.
The exact benchmarks are realized using QuSpin~\cite{weinberg2019quspin} and the finite-temperature Quantum Monte Carlo calculations using the ALPS library~\cite{albuquerque_alps_2007,bauer_alps_2011}.
The code for the simulations and the produced data are available at~\cite{repo}.

\begin{acknowledgments}
We thank R. Martinazzo, Z. Denis, L. L. Viteritti, and L. Fioroni for insightful discussions and M. Bukov for helping with the exact simulations. 
A. S. is supported by SEFRI under Grant No.\ MB22.00051 (NEQS - Neural Quantum Simulation). 
F.V. acknowledges support by the French Agence Nationale de la Recherche through the NDQM project, grant ANR-23-CE30-0018.
We acknowledge the EuroHPC Joint Undertaking for awarding this project access to the EuroHPC supercomputer LEONARDO, hosted by CINECA (Italy) and the LEONARDO consortium through the EuroHPC Development Access call EHPC-DEV-2024D10-055.
\end{acknowledgments}

\emph{Note: during the preparation of this manuscript, we became aware of a related work that has been carried on in parallel by A. Van de Walle, M. Schmitt, and A. Bohrdt, which will appear simultaneously on the preprint server. 
}

\section*{End Matter} 

\paragraph*{Appendix A: Optimization in sub-intervals --}
We have experimentally observed that converging to the minimum of $L_{[0,T]}(\theta)$ becomes harder as the final time $T$ is larger, especially for bigger system sizes, and increasing the number of basis states $M$ does not counter this sufficiently.
We therefore partition the time evolution into several sub-intervals of length $\Delta T$ and we solve the dynamics in each of them sequentially. 
At every $i$-th sub-interval, the initial condition is taken to be the wave function at time $i\Delta T$ obtained from the solution of the previous sub-interval.
In the 1D simulations, we employ $\Delta T = 0.25$ or $0.5$ depending on the specific quench, while for the 2D lattice $\Delta T = 0.2$ for all the dynamics. 
The time-integral in each sub-interval is estimated with the Simpson's $1/3$ rule using $128 + 1$ integration points in 1D with $\Delta T = 0.25$, and using $256 + 1$ points in 1D with $\Delta T = 0.5$ and in all the 2D calculations. 

\paragraph*{Appendix B: Parametrization for the coefficients --}
The time-dependent coefficients $c_i(t)$ must satisfy the initial conditions $c_i(t=0)=\delta_{i,0}$. 
To achieve this, we set $c_0(t)=1 \,\,\forall t$ and we expand the other coefficients in a truncated Fourier basis respecting the initial condition
\begin{equation}
\label{eq:coefficients}
c_i(t) = \sum_{k=1}^{N_b} \gamma_{ik}(e^{i\omega_k t}-1), \quad i > 0
\end{equation}
where $N_b$ is the number of basis functions, $\gamma_{ik}$ are variational parameters, and the frequencies $\omega_k$ are initialized to energies evenly spaced in the spectrum of $H$.
The minimum and the maximum energies of $H$ are estimated by Variational Monte Carlo.
The choice of the $\omega_k$ ensures proper coverage of the relevant dynamical time scales and is motivated also by the coarse-grained dynamics of Ref.~\cite{PhysRevB.103.024304}.
To enhance the expressivity of the ansatz, however, we allow the frequencies to be variational, so that the set of parameters $\theta$ includes $\gamma_{ik}$ and $\omega_k$ as well. 
For the 1D simulations we use $N_b$ = 64, whereas in 2D we set $N_b = 128$.

\paragraph*{Appendix C: Time-dependent linear variational method --}
For the linear ansatz~\cref{eq:ansatz} with fixed basis states $\{\ket{\phi_0}\equiv \ket{\Psi_0}, \ket{\phi_1}, \ldots, \ket{\phi_{M}}\}$, there exist optimal trajectories for the coefficients $c_i(t)$ exactly solving the Schrödinger's equation in the subspace spanned by the basis.
These are given by the equations of the time-dependent linear variational method
\begin{equation}
\label{eq:exact_coefficients}
c(t) = \exp(-i t \, \mathbb{S}^{-1} \mathbb{H}) c(0),  
\end{equation}
where $c(0) = [1, 0, \ldots, 0]$ is a $M+1$-dimensional vector setting the initial condition, $\mathbb{S}$ is the overlap/Gram matrix of the basis states, and $\mathbb{H}$ is the reduced Hamiltonian matrix in the subspace of the basis~\cite{motta2024subspace}. 
Their matrix elements are given by $\mathbb{S}_{ij} = \bra{\phi_i} \ket{\phi_j}$ and $\mathbb{H}_{ij} = \bra{\phi_i} H \ket{\phi_j}$ for $i, j = 0, \ldots, M$. 
The proof of~\cref{eq:exact_coefficients} is reported in the Supplemental Material~\cite{suppmat}.
In this work, we consider a two-step approach. 
At first, we variationally parametrize the coefficients $c_i(t)$ and we minimize the global loss with respect to the parameters of both the basis states and the coefficients. 
Then, we improve the dynamics by plugging the optimal coefficients calculated from~\cref{eq:exact_coefficients} with the optimized basis states. 
The matrix elements of $\mathbb{S}$ and $\mathbb{H}$ can be efficiently estimated (up to an irrelevant constant) through Monte Carlo sampling, as shown in the Supplemental Material~\cite{suppmat}.

\paragraph*{Appendix D: Long-time quantum dynamics --}
With the Galerkin ansatz~\cref{eq:ansatz} decoupling spatial and time degrees of freedom, it is particularly natural to gain access to the long-time dynamics of physical quantities. 
This can be done, for example, by considering the discrete Fourier transform of the optimal coefficients $c_i(t)=\sum_{k} \tilde{\gamma}_{ik} e^{i \tilde{\omega}_k t}$, where $\tilde{\omega}_k$ are the Fourier frequencies and $\tilde{\gamma}_{ik}$ the corresponding amplitudes. 
In the limit of $t\rightarrow \infty$, all the oscillating factors in the numerator and in the denominator of an expectation value have a negligible contribution with respect to the non-oscillating terms, so they can be discarded leading to 
\begin{equation}
\label{eq:thermalization}
\frac{\bra{\Psi_{\theta}(t)} O \ket{\Psi_{\theta}(t)}}{\bra{\Psi_{\theta}(t)} \ket{\Psi_{\theta}(t)}} \overset{t \rightarrow \infty}{\approx}  \frac{\sum_{ijk}\tilde{\gamma}^*_{ik} \mathbb{O}_{ij} \tilde{\gamma}_{jk}}{\sum_{ijk}\tilde{\gamma}^*_{ik} \mathbb{S}_{ij} \tilde{\gamma}_{jk}}, 
\end{equation}
where $\mathbb{O}_{ij} = \bra{\phi_i} O\ket{\phi_j}$. 
A similar expression can be found for the loss function by formulating it in terms of the $c_i(t)$, $\mathbb{S}_{ij}$, $\mathbb{H}_{ij}$ and $\mathbb{H}^{(2)}_{ij} = \bra{\phi_i} H^2 \ket{\phi_j}$. 

\begin{figure}[b]
    \centering
\includegraphics[width=1.0\linewidth]{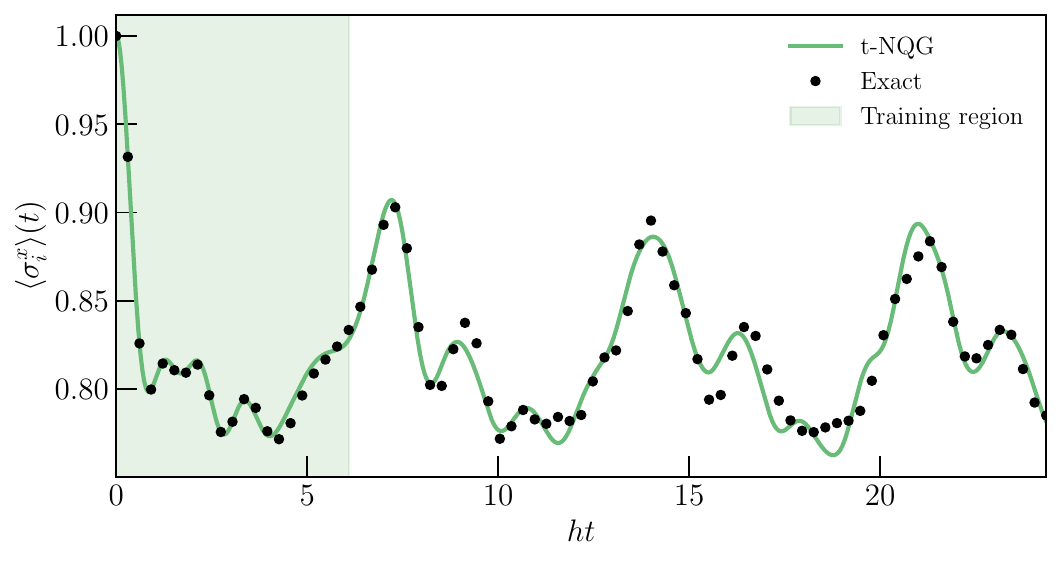}
    \caption{
    Dynamics of the transverse magnetization $\langle \sigma_i^x \rangle(t)$ extrapolated beyond the training interval $[0, T]$ with $T = 2$ (highlighted in light green). 
    The time evolution corresponds to the quench with $h = 3.044 \approx h_c^{\text{2D}}$ in the $6 \times 6$ lattice.  
    We employ $M=6$ RBM basis states, and 512 Monte Carlo samples per integration point.
    }
    \label{fig:extrapolating}
\end{figure}

\begin{figure*}[t]
    \centering
\includegraphics[width=1.0\linewidth]{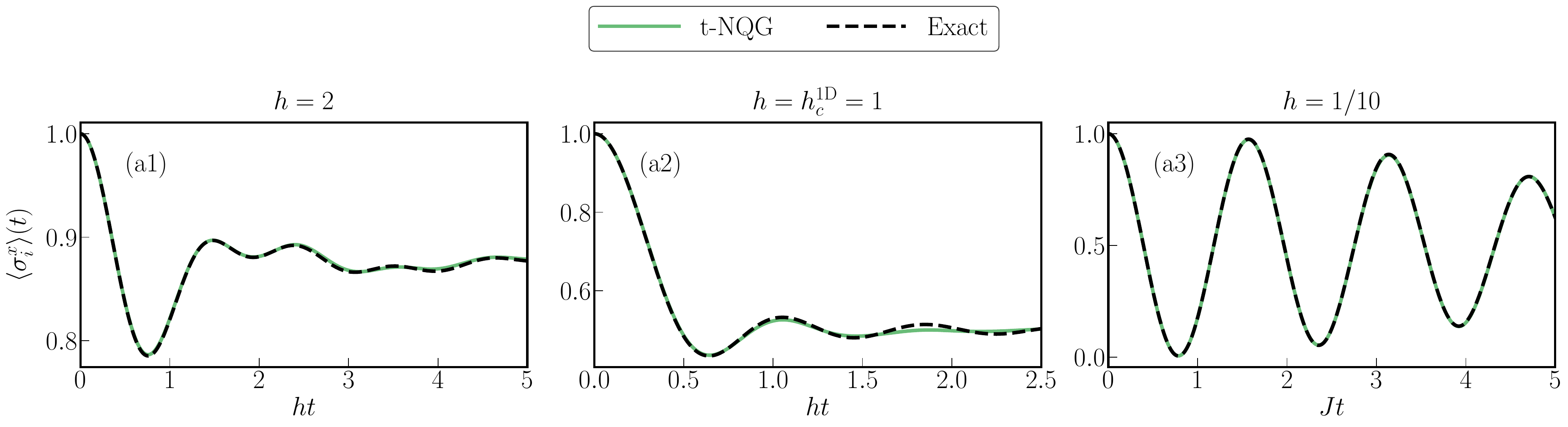}
    \caption{
    Time evolution of the transverse magnetization $\langle \sigma_i^x \rangle (t)$ following global quantum quenches in the TFI model on a 1D spin chain with $N=40$ sites.
    The system is quenched from the paramagnetically polarized initial state $|\Psi_0\rangle = \bigotimes_{i=1}^{N} |+\rangle_i$ to (a1) the paramagnetic phase at $h = 2$, (a2) the critical point at $h = h_c^{\mathrm{1D}} = 1$, and (a3) the ferromagnetic phase at $h = 1/10$.
    We employ $M = 20$ RBM basis states and 512 Monte Carlo samples per integration point.
    }
    \label{fig:1d}
\end{figure*}

\begin{figure}
    \centering
\includegraphics[width=1.0\linewidth]{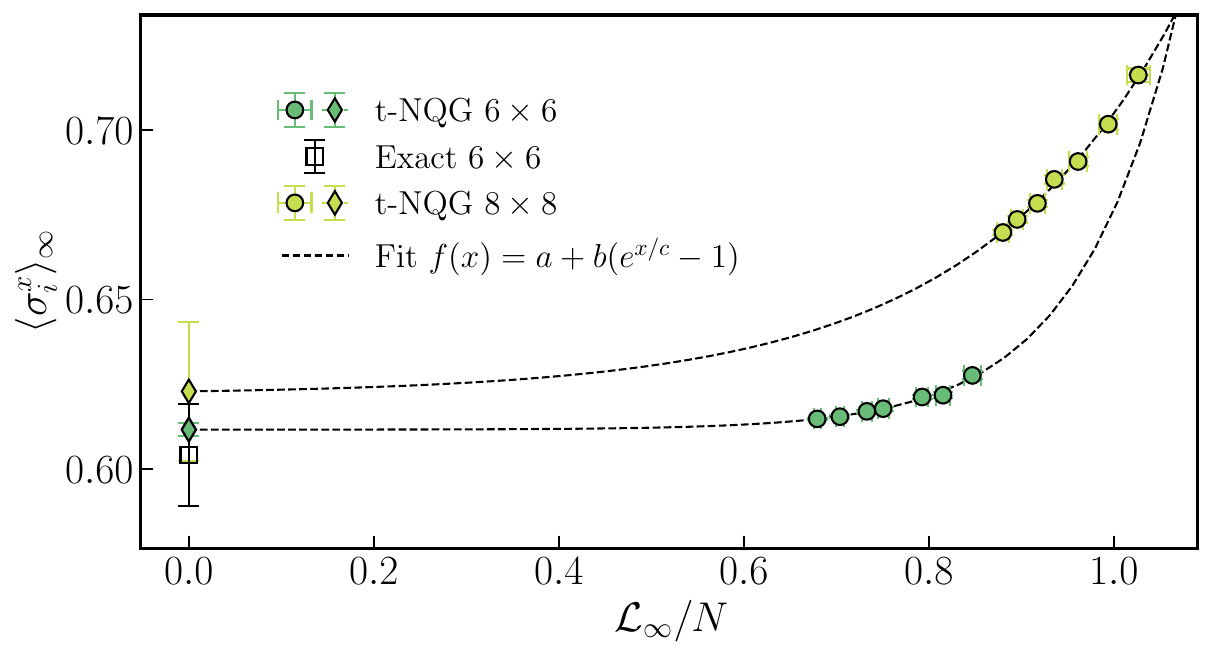}
    \caption{
    Infinite-time transverse magnetization predicted by t-NQG, $\langle \sigma_i^x \rangle_{\infty}$, as a function of the infinite-time loss $\mathcal{L}_{\infty}$ normalized by the system size $N$, for the $6 \times 6$ and $8 \times 8$ lattices after the quench at $h = 2$.
    Circle markers denote the t-NQG results, while diamonds indicate the zero-loss extrapolations.
    The uncertainty on the extrapolated value is estimated from the covariance matrix of the fit parameters.}
    \label{fig:extrapolate_zeroloss}
\end{figure}

\paragraph*{Appendix E: Extrapolation beyond the training interval --}
We demonstrate that the t-NQG method successfully extrapolates beyond the optimization interval $[0, T]$.
This is illustrated in~\cref{fig:extrapolating}, which shows that the ansatz can predict the time evolution even for $t > T$.
Although the accuracy is naturally highest within the training interval, all qualitative features of the observable’s oscillations are faithfully reproduced beyond that range.
The extrapolation is reliable provided that the training window contains the essential features of the dynamics, such as the initial transient following a quench.
Under this condition, the optimized basis states effectively encode the dominant frequencies of the time evolution, enabling faithful extrapolation beyond the training regime.
We emphasize that such extrapolation is not possible in time-stepping approaches like t-VMC~\cite{carleo_localization_2012,carleo_light-cone_2014,carleo2017solving} or its projected variant~\cite{Donatella2023Infidelity,sinibaldi2023unbiasing,nys2024ab,Gravina2024ptVMC}, which rely on sequential updates that inherently preclude extension beyond the simulated time interval.

\paragraph*{Appendix F: Dynamics in one dimension --}
As a benchmark, we also investigated the analogous quenches of in~\cref{fig:2d_together} in the 1D system. 
The exact solution is computed from fermionization of the TFI model on the one-dimensional chain with periodic boundary conditions.
The results, displayed in~\cref{fig:1d}, demonstrate that the t-NQG method accurately captures the time evolution in 1D across different quench protocols as well.

\paragraph*{Appendix G: Zero-loss extrapolation --}
To further assess the accuracy of the infinite-time calculations, we perform a zero-loss extrapolation of the observable~\cite{PhysRevB.88.060402,nomura2021helping,viteritti2023transformer_2d}.
Specifically, we carry out t-NQG simulations with varying number of basis states $M$, obtaining different values for $\langle \sigma_i^x \rangle_\infty$ and $\mathcal{L}_\infty$.
We then fit the magnetization as a function of the loss and extrapolate to the $\mathcal{L}_\infty \rightarrow 0$ limit.
An exponential fitting function is employed, as it accurately describes the numerical data. 
The results for the quench at $h = 2$ are shown in~\cref{fig:extrapolate_zeroloss}.
For the $6 \times 6$ system, the extrapolated value agrees well with the exact result, supporting the robustness of the extrapolation procedure.
For both system sizes, the data points lie well on the fitted curves.

\bibliography{bibliography}

\begin{widetext}

\section*{SUPPLEMENTAL MATERIAL}

\subsection{Loss function
\label{sec:loss}}
Here we provide the complete derivation of the loss function used in this work.
In all the following, the explicit time-dependence is omitted to simplify the notation.
In general, to satisfy the time-dependent Schrödinger's equation with a variational ansatz $\ket{\Psi_{\theta}}$ one can minimize the $L^2$ distance between the two sides of the equation
\begin{equation}
\label{eq:L2}
    |\!| \ket*{\dot{\Psi}_{\theta}} + i H \ket{\Psi_{\theta}}|\!|^2.
\end{equation}

For physical applications, however, the loss function above is unsatisfactory, since it does not incorporate the geometry of quantum states. Specifically, a physically robust loss function must be invariant under two fundamental transformations: arbitrary (possibly time-dependent) changes in the normalization and global phase rotations of the state $\ket{\Psi_{\theta}}$. 
To make the loss~\cref{eq:L2} invariant under norm changes, it is enough to consider the distance between the normalized states, namely 
\begin{equation}
\label{eq:loss_norm}
        \mathcal{L}^{\prime}(\ket{\Psi_{\theta}}) =  \bigg|\!\bigg| \frac{d}{dt}\bigg(\frac{\ket*{\Psi_{\theta}}}{\sqrt{\bra{
        \Psi_{\theta}}\ket{\Psi_{\theta}}}}\bigg) + i H \frac{\ket*{\Psi_{\theta}}}{\sqrt{\bra{
        \Psi_{\theta}}\ket{\Psi_{\theta}}}} \bigg|\!\bigg|^2.
\end{equation}

To guarantee the invariance under the phase variation, we consider how~\cref{eq:loss_norm} is modified after the transformation $\ket{\Psi_{\theta}} \rightarrow e^{i \phi} \ket{\Psi_{\theta}}$, where $\phi = \phi(t) \in \mathbb{R}$. 
We obtain
\begin{equation}
\begin{split}
\label{eq:loss_trasformed}
    \mathcal{L}^{\prime}(e^{i \phi}\ket{\Psi_{\theta}}) &= 
    \bigg|\!\bigg|  \frac{d}{dt} \bigg(\frac{e^{i\phi} \ket{\Psi_{\theta}}}{\sqrt{\bra{\Psi_{\theta}} \ket{\Psi_{\theta}}}}\bigg) + i e^{i \phi} H \frac{\ket{\Psi_{\theta}}}{\sqrt{\bra{\Psi_{\theta}} \ket{\Psi_{\theta}}}} \bigg|\!\bigg| ^2 = \bigg|\!\bigg| \frac{d}{dt} \bigg(\frac{ \ket{\Psi_{\theta}}}{\sqrt{\bra{\Psi_{\theta}} \ket{\Psi_{\theta}}}}\bigg)  + i   H \frac{\ket{\Psi_{\theta}}}{\sqrt{\bra{\Psi_{\theta}} \ket{\Psi_{\theta}}}}  +  i \dot{\phi}\frac{ \ket{\Psi_{\theta}}}{\sqrt{\bra{\Psi_{\theta}} \ket{\Psi_{\theta}}}}\bigg|\!\bigg| ^2 = \\
    &= \bra{\epsilon_{\theta}} \ket{\epsilon_{\theta}} + 2 \Re\bigg(i \frac{\bra{\epsilon_{\theta}} \ket{\Psi_{\theta}}}{\sqrt{\bra{\Psi_{\theta}} \ket{\Psi_{\theta}}}}\bigg) \dot{\phi} + \dot{\phi}^2, 
\end{split}
\end{equation}
where we denote $\ket{\epsilon_{\theta}} \equiv \dfrac{d}{dt} \bigg(\dfrac{ \ket{\Psi_{\theta}}}{\sqrt{\bra{\Psi_{\theta}} \ket{\Psi_{\theta}}}}\bigg)  + i   H \dfrac{\ket{\Psi_{\theta}}}{\sqrt{\bra{\Psi_{\theta}} \ket{\Psi_{\theta}}}}$. 
We note that~\cref{eq:loss_trasformed} depends only on $\dot{\phi}$ and not on $\phi$. 
We can impose that the loss is invariant under phase variations by choosing the $\dot{\phi}$ which minimizes $\mathcal{L}^{\prime}(e^{i \phi}\ket{\Psi_{\theta}})$, in the same way as done in the time-dependent variational principle~\cite{Yuan2019theoryofvariational}.
Thus, by putting $\partial_{\dot{\phi}} \mathcal{L}^{\prime}(e^{i \phi} \ket{\Psi_{\theta}}) = 0$ we get the optimal phase velocity 
\begin{equation}
\label{eq:optimal_velocity}
    \dot{\phi} = \Im\bigg(\frac{\bra*{\dot{\Psi}_{\theta}}\ket{\Psi_{\theta}}}{\bra{\Psi_{\theta}}\ket{\Psi_{\theta}}}\bigg) - \langle H \rangle,
\end{equation}
where $\langle H \rangle = \bra{\Psi_{\theta}} H \ket{\Psi_{\theta}} / \bra{\Psi_{\theta}} \ket{\Psi_{\theta}}$.
Plugging~\cref{eq:optimal_velocity} into~\cref{eq:loss_trasformed} and rewriting yields the fully norm and phase invariant expression used in the main text 
\begin{equation}
\label{eq:loss_derived}
        \mathcal{L}(\ket{\Psi_{\theta}}) = \bigg|\!\bigg| \bigg(1 - \frac{\ket{\Psi_{\theta}}\bra{\Psi_{\theta}}}{\bra{\Psi_{\theta}}\ket{\Psi_{\theta}}}\bigg)\bigg(\frac{\ket*{\dot{\Psi}_{\theta}}}{\sqrt{\bra{\Psi_{\theta}}\ket{\Psi_{\theta}}}} + i H  \frac{\ket{\Psi_{\theta}}}{\sqrt{\bra{\Psi_{\theta}}\ket{\Psi_{\theta}}}} \bigg)\bigg|\!\bigg| ^2.
\end{equation}

\subsection{Monte Carlo evaluation of the loss function \label{sec:loss_mc}}
The loss function $\mathcal{L}(\ket{\Psi_{\theta}})$ can be efficiently evaluated using Monte Carlo sampling. Here we provide the complete derivation of the stochastic estimators.  
We start by introducing the operator $L$ through its action on $\ket{\Psi_{\theta}}$
\begin{equation}
\label{eq:L_operator}
L|\Psi_{\theta}\rangle = |\dot{\Psi}_{\theta}\rangle + iH|\Psi_{\theta}\rangle.
\end{equation}

We remark that $L$ is not Hermitian in general.
The loss $\mathcal{L}(\ket{\Psi_{\theta}})$ can be compactly written as
\begin{equation}
\label{eq:loss_operator}
\mathcal{L}(\ket{\Psi_{\theta}}) = \frac{\langle\Psi_{\theta}|L^\dagger L|\Psi_{\theta}\rangle}{\langle\Psi_{\theta}|\Psi_{\theta}\rangle} - \frac{\langle\Psi_{\theta}|L^\dagger|\Psi_{\theta}\rangle\langle\Psi_{\theta}|L|\Psi_{\theta}\rangle}{\langle\Psi_{\theta}|\Psi_{\theta}\rangle \langle\Psi_{\theta}|\Psi_{\theta}\rangle}. 
\end{equation}

The previous expression corresponds to the quantum variance of the operator $L$.
By introducing the completeness relation of a basis $\{\ket{\sigma}\}$ of the Hilbert space,~\cref{eq:loss_operator} can be evaluated as the statistical variance of the local estimator of $L$
\begin{equation}
\mathcal{L}(\ket{\Psi_{\theta}}) = \mathbb{E}_{|\Psi_{\theta}(\sigma)|^2}[|L_\text{loc}(\sigma)|^2] - |\mathbb{E}_{|\Psi_{\theta}(\sigma)|^2}[L_\text{loc}(\sigma)]|^2 = \mathbb{E}_{|\Psi_{\theta}(\sigma)|^2} [|L_\text{loc}(\sigma) -\mathbb{E}_{|\Psi_{\theta}(\sigma)|^2}[L_\text{loc}(\sigma)]|^2], 
\end{equation}
where the local estimator is
\begin{equation}
L_\text{loc}(\sigma) = \frac{\bra{\sigma} L \ket{\Psi_{\theta}}}{\bra{\sigma}\ket{\Psi_{\theta}}} = O_t(\sigma) + iE_\text{loc}(\sigma),
\end{equation}
where $O_t(\sigma) = \partial_t \log \Psi_{\theta}(\sigma) = \dot{\Psi}_{\theta}(\sigma)/\Psi_{\theta}(\sigma)$ is the logarithmic time derivative and $E_\text{loc}(\sigma) = \langle\sigma|H|\Psi_{\theta}\rangle/\Psi_{\theta}(\sigma)$ is the local energy.

\subsection{Equations of the time-dependent linear variational method}
To derive the formula for the coefficients from the time-dependent linear variational method, we first write the Schrödinger's equation for the linear ansatz obtaining
\begin{equation}
\label{eq:schrodinger_projected}
    \sum_{j} \dot{c}_j(t) \ket{\phi_j} = -i \sum_j c_j(t) H\ket{\phi_j}.
\end{equation}

Then, we search for a solution of~\cref{eq:schrodinger_projected} in the subspace spanned by $\{\ket{\phi_i}\}$ by projecting it onto each basis state, leading to the set of equations
\begin{equation}
\label{eq:coeff_ode}
    \sum_j \mathbb{S}_{ij} \dot{c}_j(t) = -i \sum_j  \mathbb{H}_{ij} c_j(t) \hspace{0.5cm} \forall \,\, i, 
\end{equation}
where $\mathbb{S}_{ij} = \bra{\phi_i} \ket{\phi_j}$ and $\mathbb{H}_{ij} = \bra{\phi_i}H \ket{\phi_j}$.
The solution of the system of ordinary differential equations~\cref{eq:coeff_ode} is known and corresponds to: 
\begin{equation}
\label{eq:optimal_coefficients}
    c(t) = \exp(-i t \, \mathbb{S}^{-1} \mathbb{H}) c(0), 
\end{equation}
with $c(0)$ setting the initial condition.

\subsection{Monte Carlo estimation of the matrix elements \label{sec:matrix_mc}}
For all the computations in the subspace spanned by the basis states, it is enough to know the overlap matrix $\mathbb{S}$ and the matrix representation of any observable $O$ in the basis, say $\mathbb{O}$, up to a common constant $k$.
This observation is essential to be able to estimate their entries with Monte Carlo sampling, since in general the states $\ket{\phi_i}$ are not normalized. 
Therefore, we can introduce an arbitrary probability distribution $\Pi(\sigma)$ that can be used to estimate $\mathbb{S}_{ij} / k$ and $\mathbb{O}_{ij} / k$ where $k$ corresponds to the normalization of $\Pi$. 
Indeed, we can write
\begin{equation}
\label{eq:matrix_mc_explicit}
\begin{split}
\frac{\mathbb{S}_{ij}}{\sum_{\sigma^{\prime}} \Pi(\sigma^{\prime})} &= \frac{\sum_{\sigma^{}} \Pi(\sigma)}{\sum_{\sigma^{\prime}}\Pi(\sigma^{\prime})}\bigg[\frac{\phi^*_i(\sigma)\phi_j(\sigma)}{\Pi(\sigma)}\bigg] = \mathbb{E}_{\Pi(\sigma)} \bigg[\frac{\phi^*_i(\sigma)\phi_j(\sigma)}{\Pi(\sigma)}\bigg],\\
\frac{\mathbb{O}_{ij}}{\sum_{\sigma^{\prime}} \Pi(\sigma^{\prime})} &= \frac{1}{2}\frac{\sum_{\sigma} \Pi(\sigma)}{\sum_{\sigma^{\prime}}\Pi(\sigma^{\prime})}\bigg[\frac{\phi^*_i(\sigma)\bra{\sigma} O \ket{\phi_j}}{\Pi(\sigma)} + \frac{\bra{\phi_i}O\ket{\sigma}\phi_j(\sigma)}{\Pi(\sigma)}\bigg] = \frac{1}{2}\mathbb{E}_{\Pi(\sigma)} \bigg[\frac{\phi^*_i(\sigma)\bra{\sigma} O \ket{\phi_j}}{\Pi(\sigma)} + \frac{\bra{\phi_i}O\ket{\sigma}\phi_j(\sigma)}{\Pi(\sigma)}\bigg].
\end{split}
\end{equation}

We note that for the observable matrix we employ a symmetrized estimator. 
Since we need a distribution $\Pi$ with support over all the basis states to accurately estimate all the expectation values in~\cref{eq:matrix_mc_explicit}, we consider $\Pi(\sigma) = \sum_{i=0}^{M} |\phi_i(\sigma)|^2$.

\subsection{Optimal number of basis states}
Here we present a proof showing that with the linear Galerkin ansatz it is possible to keep the error of the variational dynamics below a small finite value by increasing the number of basis states $M$ polynomially in time $t$ and in the system size $N$.
For the specific purpose of the calculation, we consider the basis states to be the ones of the coarse-grained (CG) approach to quantum dynamics~\cite{PhysRevB.103.024304}, but this analysis is general and can be extended to any enough expressive set of basis states such as Lanczos or Chebyshev vectors~\cite{motta2024subspace}. 

In the CG approach, the approximate time-evolved state is constructed as $| \Psi^{ \text{CG}}(t) \rangle = \sum_{i=0}^{M-1} e^{-it \lambda_i}|w_i\rangle$, where $|w_i\rangle$ are the CG basis states and $\lambda_i = \frac{\bra{w_i} H \ket{w_i}}{\bra{w_i} \ket{w_i}}$.
The CG basis states $|w_i \rangle$ are obtained by minimizing
\begin{equation}
\label{eq:loss_cg}
    \sum_{i=0}^{M-1} \langle w_i |(H - \mu_i)^2|w_i\rangle - 2 \text{Re}\bigg\{ \langle \lambda | \bigg(| w_i \rangle - \ket{\Psi_0}\bigg)\bigg\},
\end{equation}
where $\mu_i$ are fixed energies uniformly distributed in the spectrum of $H$ such that $\mu_0 \leq E_{\text{min}}$ and $\mu_{M-1} \geq E_{\text{max}}$, where
$E_{\text{min}}, E_{\text{max}}$ are the minimum and the maximum eigenvalues of $H$ respectively. 
The energy separation between the $\mu_i$ is indicated as $\epsilon$.
$|\lambda \rangle$ is the Lagrange multiplier for the constraint on the initial condition $|\Psi_0 \rangle = \sum_{i=0}^{M-1} |w_i \rangle$.  
In practice, the states $|w_i\rangle$ capture the components of the initial state $| \Psi_0 \rangle$ on the energy eigenstates with energies closest to $\mu_i$. 
The minimum of~\cref{eq:loss_cg} is found by setting its gradient with respect to $\bra{w_i}$ to zero and by expressing $\ket{\lambda}$ in terms of $\ket{\Psi_0}$ from the constraint condition.
The solution can be written as $|w_i \rangle = W_i | \Psi_0 \rangle$ with the weighting operator $W_i$ defined as $W_i  =  \frac{(H -\mu_i)^{-2}}{\sum_{j=0}^{M-1} (H -\mu_j)^{-2}}$.

As shown in~\cite{PhysRevB.103.024304}, for small times $t$ the error can be approximated as $\delta(t) \equiv |\!|e^{-i H t}|\Psi_0\rangle - |\Psi^{\text{CG}}(t)\rangle |\!| \approx \sqrt{\sigma^2} t$, where $\sigma^2 = \frac{\sum_{i=0}^{M-1} \bra{w_i} (H - \lambda_i)^2 \ket{w_i}}{\sum_{i=0}^{M-1} \bra{w_i} \ket{w_i}}$.
Now, assuming that $\frac{\bra{w_i}(H - \lambda_i)^2 \ket{w_i}}{\bra{w_i} \ket{w_i}} \sim \epsilon^2$~\cite{PhysRevB.103.024304}, which is true when $M$ is large enough, we obtain:
\begin{equation}
   \delta(t) \sim \epsilon \, t \approx \bigg(\frac{E_{\mathrm{max}}-E_{\mathrm{min}}}{M}\bigg) t = O\bigg(\frac{N t}{M}\bigg),
\end{equation}
in the case of a physical Hamiltonian with an extensive energy spectrum.
This implies that, to keep the error at time $t$ below $\bar{\delta}$, the number of basis states must scale as $M \sim \frac{N t}{\bar{\delta}}$.

\subsection{Error with the exact dynamics}
The time evolution of the linear ansatz with fixed basis states and the optimal coefficients from the time-dependent linear variational method can be written in terms of the projected Hamiltonian $H_Q = QHQ$ as $|\Psi_{\theta}(t) \rangle = e^{-i t H_Q} \ket{\Psi_0}$, where $Q = \sum_{i, j=0}^{M} (\mathbb{S}^{-1})_{ij} | \phi_i \rangle \langle \phi_j |$ is the orthogonal projector onto the subspace spanned by the basis. 
For simplicity, we consider the initial state $\ket{\Psi_0}$ to have unit norm such that $\ket{\Psi_{\theta}(t)}$ remains normalized at all times, but the calculation is completely general.
Under this condition, our local-in-time loss function $\mathcal{L}(t) \equiv \mathcal{L}(\ket{\Psi_{\theta}(t)})$ becomes
\begin{equation}
\label{eq:residual}
\mathcal{L}(t) = |\!| P_{\perp \ket{\Psi_{\theta}(t)}}(\ket*{\dot{\Psi}_{\theta}(t)} + i H \ket{\Psi_{\theta}(t)})|\!|^2 = |\!|\ket{r(t)}|\!|^2,
\end{equation}
where the residual state is defined as $|r(t)\rangle =  - i H \ket{\Psi_{\theta}(t)} - \ket*{\dot{\Psi}_{\theta}(t)} = -i(1-Q)H| \Psi_{\theta}(t) \rangle $.
The last equality in~\cref{eq:residual} is non-trivial and comes from the fact that $P_{\perp \ket{\Psi_{\theta}(t)}}\ket{r(t)} = \ket{r(t)}$ since $(1 - \ket{\Psi_{\theta}(t)}\bra{\Psi_{\theta}(t)})(1-Q) = 1-Q$.
We remark that the loss can also be written in terms of the matrices in the basis subspace as $\mathcal{L}(t) = |\!|\ket{r(t)}|\!|^2 = c^\dagger(t) \Sigma  c(t)$, where $c(t)$ correspond to the optimal coefficients~\cref{eq:optimal_coefficients} and $\Sigma =  \mathbb{H}^{(2)} - \mathbb{H} \mathbb{S}^{-1} \mathbb{H}$ with $\mathbb{H}^{(2)}_{ij} = \bra{\phi_i} H^2 \ket{\phi_j}$.

The error with respect to the exact evolution $ |\!|\ket{\epsilon(t)}|\!| = |\!||\Psi(t) \rangle - |\Psi_{\theta}(t) \rangle|\!|$ with $\ket{\Psi(t)} = e^{-i t H} \ket{\Psi_0}$ can be bounded by the global loss $L_{[0,t]} = \frac{1}{t} \int_0^t dt^\prime \mathcal{L}(t^\prime)$.
Indeed, we can write:
\begin{equation}
    | \dot{\epsilon} (t) \rangle = | \dot{\Psi}(t) \rangle - | \dot{\Psi}_{\theta}(t) \rangle = \ket{r(t)} - i H \ket{\epsilon(t)},
\end{equation}
from which we recover the following differential equation for $\ket{\epsilon(t)}$: 
\begin{equation}
    \bigg(\frac{d}{dt} + i H \bigg)\ket{\epsilon(t)} =  \ket{r(t)}.
\end{equation}

The analytical solution of the previous equation can be written using the Green's function formalism, obtaining:
\begin{equation}
    | \epsilon(t) \rangle =  e^{-i H t} \int_0^t dt^\prime e^{i H t^\prime} | r(t^\prime) \rangle.
\end{equation}

Therefore, we can bound the error with the exact dynamics as:
\begin{equation}
\label{eq:state_bound}
    |\!| | \epsilon(t) \rangle |\!| \leq \int_0^t dt^\prime |\!|  | r(t^\prime) \rangle |\!| \leq \sqrt{t \int_0^t dt^\prime |\!|  | r(t^\prime) \rangle |\!|^2}  = t \sqrt{L_{[0,t]}}.
\end{equation}

From this, it is also possible to bound the error on the expectation value of any observable $O$ at each time as: 
\begin{equation}
\label{eq:observable_bound}
\begin{split}
|\delta O (t)| &= |\langle O \rangle_{\ket{\Psi(t)}} -\langle O \rangle_{\ket{\Psi_{\theta}(t)}}| = \\
&= |2 \text{Re}\langle \epsilon(t) | O | \Psi(t) \rangle - \langle \epsilon(t) | O | \epsilon(t) \rangle| \\
&\leq |\!|O|\!|_2 (2|\!|\ket{\epsilon(t)}|\!| + |\!|\ket{\epsilon(t)}|\!|^2) \\ 
&\leq |\!|O|\!|_2(2 t \sqrt{L_{[0,t]}} + t^2 L_{[0,t]}),
\end{split}
\end{equation}
where $|\!|O|\!|_2$ indicates the operator 2-norm of $O$.
The bounds in~\cref{eq:state_bound,eq:observable_bound} are derived without any assumption on the time $t$, although we remark that for large $t$ they may become lax in practice.
Moreover, we observe that for local observables the actual accuracy is typically much higher than what the bound~\cref{eq:observable_bound} alone would suggest.
This behavior is evident in the 2D simulations shown in Fig. 2 of the main text: while the loss values in the inset would imply rather large deviations via~\cref{eq:observable_bound}, the agreement with the exact dynamics is in fact significantly better.
A similar behavior is expected to hold for the observable at infinite-time when considering the infinite-time loss function $\mathcal{L}_{\infty} \equiv \lim_{t \rightarrow \infty} \mathcal{L}(|\Psi_{\theta}(t)\rangle)$ as a measure of accuracy.

\subsection{Number of basis states and final time in the optimization}
\begin{figure}[b]
    \centering
\includegraphics[width=1.0\linewidth]{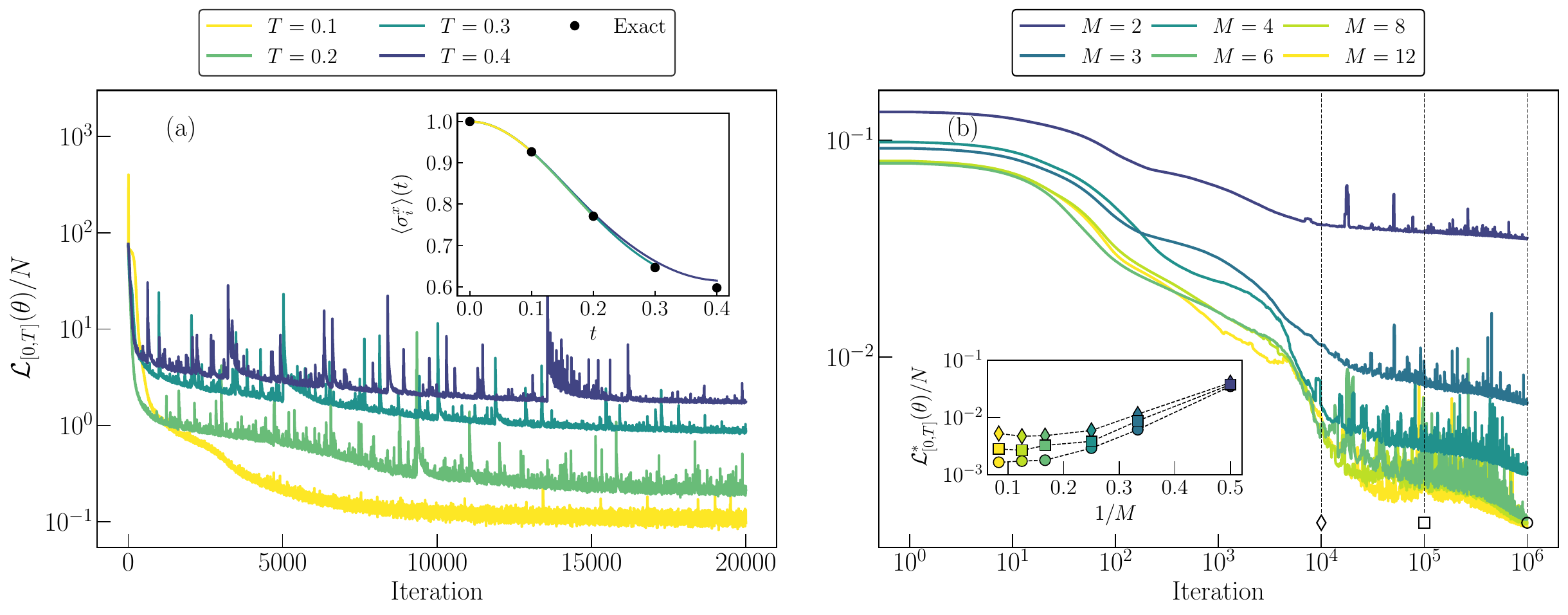}
    \caption{
    (a) Learning curves of the loss function $\mathcal{L}_{[0, T]}(\theta)$, normalized by the system size $N$, for different final times $T$.
    The inset displays the corresponding dynamics of the transverse magnetization $\langle \sigma_i^x \rangle (t)$, compared with the exact simulation.
    We use $M = 6$ basis states.
    (b) Learning curves of the loss function $\mathcal{L}_{[0, T]}(\theta)$, normalized by the system size $N$, for different number of basis states $M$. 
    The inset displays the converged loss value $\mathcal{L}^*_{[0, T]}(\theta)$, computed as the mean over the last $10^3$ iterations, as a function of $1/M$ at various stages of the optimization (see dashed vertical lines) indicated with different markers.
    The final time considered is $T=0.2$.
    For both (a) and (b), the time evolution corresponds to the quench with $h = 2$ in the $6 \times 6$ lattice.
    The basis states are RBMs, and 512 Monte Carlo samples per integration point are employed.
     }
    \label{fig:scalings}
\end{figure}

We analyze how the accuracy of the method scales, in practice, with the number of basis states $M$ and the final time $T$.
For a larger $T$, accurately simulating the dynamics typically requires representing a broader set of distinct time-evolved states.
Consequently, for a fixed number of basis states $M$, the minimum achievable loss is expected to increase with $T$, resulting in less accurate dynamics.
This trend is precisely illustrated in~\cref{fig:scalings}(a).
Conversely, choosing $T$ too small increases the computational cost, as more concatenations are required to reach the same total evolution time (see Appendix A of the main text).
To ensure scalability while preserving accuracy, one must therefore select a final time that optimally balances computational efficiency and fidelity of the dynamics.

We also investigate, for a fixed $T$, how increasing $M$ improves the simulation by enhancing the expressivity of the ansatz.
As shown in~\cref{fig:scalings}(b), the learning curves systematically reach lower values as $M$ increases.
The inset further indicates that the converged loss decreases approximately as $1/M$, even across different stages of the training.
However, beyond a critical value of $M$, the improvement saturates, as evidenced by the flattening of the optimal loss for $M \geq 6$.
We attribute this saturation primarily to the increasing difficulty of simultaneously optimizing a larger number of basis states, particularly when using first-order optimizers such as Adam, which tend to struggle to converge in high-dimensional parameter spaces.
\end{widetext}

\end{document}